# Lateral heterostructures of two-dimensional materials by electron-beam induced stitching

Andreas Winter[a,1], Antony George[a,1], Christof Neumann[a], Zian Tang[a], Michael J. Mohn[b], Johannes Biskupek[b], Nirul Masurkar[c], Leela Mohana Reddy Arava[c], Thomas Weimann[d], Uwe Hübner[e], Ute Kaiser[b], Andrey Turchanin[a, f]*

[a] *Institute of Physical Chemistry, Friedrich Schiller University Jena, 07743 Jena, Germany*
[b] *Electron Microscopy Group of Materials Science, Ulm University, 89081 Ulm, Germany*
[c] *Department of Mechanical Engineering, Wayne State University, 48202 Detroit, USA*
[d] *Physikalisch-Technische Bundesanstalt, 38116 Braunschweig, Germany*
[e] *Leibniz Institute of Photonic Technology, 07745 Jena, Germany*
[f] *Jena Center for Soft Matter (JCSM), 07743 Jena, Germany*

*Corresponding author

E-mail address: andrey.turchanin@uni-jena.de

[1] These authors contributed equally to the work.

**Abstract**

We present a novel methodology to synthesize two-dimensional (2D) lateral heterostructures of graphene and $MoS_2$ sheets with molecular carbon nanomembranes (CNMs), which is based on electron beam induced stitching. Monolayers of graphene and $MoS_2$ were grown by chemical vapor deposition (CVD) on copper and $SiO_2$ substrates, respectively, transferred onto gold/mica substrates and patterned by electron beam lithography or photolithography. Self-assembled monolayers (SAMs) of aromatic thiols were grown on the gold film in the areas where the 2D materials were not present. An irradiation with a low energy electron beam was employed to convert the SAMs into CNMs and simultaneously stitching the CNM edges to the edges of graphene and $MoS_2$, therewith forming a heterogeneous but continuous film composed of two different materials. The formed lateral heterostructures possess a high mechanical stability, enabling their transfer from the gold substrate onto target substrates and even the preparation as freestanding sheets. We characterized the individual steps of this synthesis and the structure of the final heterostructures by complementary analytical techniques including optical microscopy, Raman spectroscopy, atomic force microscopy (AFM), helium ion microscopy (HIM), X-ray photoelectron spectroscopy (XPS) and high-resolution transmission electron microscopy (HRTEM) and find that they possess nearly atomically sharp boundaries.

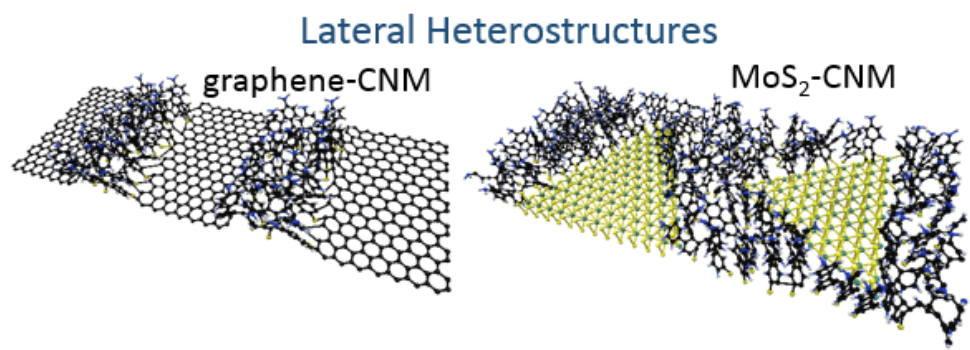

# 1. Introduction

Recent developments in the synthesis and applications of atomically or molecular thin two-dimensional (2D) materials, such as graphene, hexagonal boron nitride (hBN), transition metal dichalcogenides (TMDs) or molecular nanosheets (see e.g. [1-4]), pave the way towards their use in ultrathin integrated circuitry, required for novel electronic [5], optoelectronic [6] and energy storage devices [7]. For such implementations, it is essential to develop methods for the integration of various 2D materials with metallic, semiconducting and insulating properties into complex *in-plane* architectures, i.e. *lateral heterostructures*, with precise spatial control and sufficient mechanical and thermal stability. In comparison to the *vertical heterostructures* of 2D materials [8-10], which can be assembled by simple mechanical stacking of the individual layers, much less work has been reported on lateral heterostructures so far [10]. Such a difference is most likely due to the necessity to employ more sophisticated synthetic approaches to induce stitching of the individual nanosheets within a plane.

One of the first reports on the growth of lateral heterostructures of 2D materials demonstrated the synthesis of graphene and hBN (graphene-hBN) lateral heterostructures by two-step chemical vapor deposition (CVD) on Ru(0001) [11]. In this method, graphene islands were initially grown on the single crystal surface using ethylene as a precursor and in the second CVD step the interisland area was filled with hBN employing borazine for the growth. The obtained graphene-hBN heterostructures were composed of nanometer sized and randomly oriented domains of both 2D materials. To enable the implementation of graphene-hBN heterostructures in devices, a methodology to produce heterostructures with controlled domain sizes was proposed soon after [12, 13]. In this approach, a large area graphene or hBN layer is grown on a metal foil in the first CVD step, then this layer is patterned in a desirable shape using conventional lithography methods and finally the second CVD step is performed to grow the heterostructure. CVD methodology employing a single step [14] or a modulated [15] deposition of the precursors was demonstrated to grow randomly oriented lateral heterostructures of different transition metal dichalcogenides including $MoS_2$, $WS_2$ and $WSe_2$. In this way, it was possible to synthesize atomically thin p-n junctions and to study their performance in electronic and optoelectronic devices [14, 15].

A prerequisite for the CVD based synthesis of 2D lateral heterostructures is a crystallographic similarity of the corresponding material pairs, as it is the case for hBN-graphene or TMD1-TMD2 heterostructures. In case of the material pairs with very different crystallographic structure, like e.g. graphene and TMDs or hBN and TMDs, their CVD based stitching seems to be unfeasible and therefore new alternative approaches have to be developed for the synthesis. Here we present a novel method to stitch 2D sheets of various materials to form lateral heterostructures by electron irradiation induced chemical reactions. We demonstrate the fabrication and characterization of the lateral heterostructures made of graphene and carbon nanomembranes (graphene-CNM) as well as $MoS_2$ and carbon nanomembranes ($MoS_2$-CNM) with well-defined domains of either of these 2D materials (see Fig. 1). CNMs are dielectric molecular nanosheets synthesized by electron irradiation induced crosslinking of aromatic self-assembled monolayers (SAMs) [16]. CNMs can be produced on a large scale from various molecular precursors, which enables the tuning of their functional properties: thickness, chemical composition, mechanical properties, etc. [17-19]. Combining CNMs with graphene, we produce heterostructures consisting of insulating and metallic domains and using the combination with $MoS_2$, the heterostructures consist of semiconductor and insulator domains. By employing conventional lithographic techniques, it is possible to produce heterostructures with complex geometries and with nearly atomically sharp grain boundaries. The lateral resolution corresponds to the resolution of the applied lithography. The mechanical robustness of graphene-CNM and $MoS_2$-CNM lateral heterostructures enables their transfer onto arbitrary substrates as well as the preparation of freestanding sheets. We characterized properties of the produced heterostructures in detail, employing a number of complementary spectroscopy and microscopy techniques including optical microscopy (OM), scanning electron microscopy (SEM), helium ion microscopy (HIM), high-resolution transmission electron microscopy (HRTEM), atomic force microscopy (AFM), Raman and X-ray photoelectron spectroscopy (XPS).

## 2. Experimental section

*2.1 Sample preparation*

*2.1.1 Growth of graphene*

Graphene samples were grown by CVD on copper foils [20, 21] (25 μm thickness, 99.8%, Alpha Aesar) in a tube furnace (Gero F40-200, base pressure $1\times10^{-3}$ mbar). First, the copper foils were cleaned in acetic acid (97%, 15 min), rinsed with isopropanol and blown dry in an argon stream. Then they were introduced to a tube furnace and annealed for 3 hours at 1015 °C in a hydrogen flow of 50 cm³/min in order to remove the oxide layer from the substrate. Subsequently, a $CH_4/H_2$ mixture with 70 cm³/min $CH_4$ (purity 4.5) and 10 cm³/min $H_2$ (purity 5.3) was introduced to the furnace for 15 minutes. After the growth, the substrates were cooled down fast by moving the heating zone of the oven away from the sample region while keeping the flow of the precursor gases constant. This procedure results in the growth of a continuous single layer graphene film on a copper foil with some inclusions of graphene double layers, Fig. S1a.

*2.1.2 Growth of MoS$_2$*

Monolayers of MoS$_2$ single crystals were grown by a CVD process [22]. Silicon substrates with a thermally grown SiO$_2$ layer of 300 nm were used as the growth substrate (Siltronix, roughness 0.3 nm RMS). The growth was carried out in a two-zone tube furnace with a tube diameter of 55 mm. The substrates were cleaned initially by ultrasonication in acetone for 5 minutes followed by washing in isopropanol and blown dry with argon. An alumina boat containing 200 mg of sulfur powder (99.98%, Sigma Aldrich) was placed in the center of the first zone of the tube furnace. The substrates were placed with the SiO$_2$ facing down on top of an alumina boat containing ~1 μg MoO$_3$ powder (99.97%, Sigma Aldrich) and loaded to the center of the second zone of the furnace. Then the quartz tube was evacuated to $5\times10^{-2}$ mbar pressure and refilled with argon. The growth was carried out at atmospheric pressure under an argon flow of 50 cm³/min. The second zone containing the MoO$_3$ and the substrates were heated to the growth temperature of 750 °C at a rate of 40 °C/min and held at that temperature for 15 minutes. The sulfur temperature was adjusted to reach 200 °C when the second zone reaches 750 °C. After the growth, the furnace was turned off and allowed to cool down under an argon flow of 50 cm³/min until 350 °C were reached. Then the tube furnace was opened to rapidly cool down to room temperature (RT). This procedures result in the growth of mainly monolayer MoS$_2$ crystals of triangular shape with a typical size of about 100 μm, Fig. S1b

*2.1.3 Transfer of graphene and MoS$_2$ monolayers*

We have employed a poly(methyl methacrylate) (PMMA) assisted transfer protocol to transfer graphene and MoS$_2$ [23]. To transfer graphene, first a PMMA layer (100 nm, 50 kDa, All-Resist, AR-P 671.04) was spin coated onto the graphene on copper foil and hardened for 10 min at 90 °C. Subsequently, a thicker layer of PMMA (200 nm, 950 kDa, All-Resist, AR-P 679.04) was spin coated on top of the first one and hardened for 10 min at 90 °C. Then the copper foil was kept floating on top of a bath containing ammonium persulfate solution (Sigma Aldrich, 2.5%, 3 h) to etch the copper foil and release the graphene. Subsequently, the graphene supported with PMMA was washed several times in ultrapure water (18.2 MΩcm, Membrapure) to remove any residual etching solution and placed on top of a gold/mica substrate (300 nm, Georg Albert PVD). After baking at 90 °C for 10 min., the samples were immersed in acetone for 2 hours to remove the stabilizing PMMA layer.

To transfer $MoS_2$, a PMMA layer of 200 nm (950kDa, All-Resist, AR-P 679.04) was spin coated onto the $SiO_2$ substrate with CVD grown $MoS_2$ crystals and hardened for 10 min at 90 °C. Then the substrate was kept floating on top of a bath of KOH solution to etch away the $SiO_2$ layer and to release the $MoS_2$ crystals supported by PMMA followed by washing several times with ultrapure water (18.2 MΩcm, Membrapure) to remove any residual KOH. Then the PMMA supported $MoS_2$ was placed on a gold/mica substrate (300 nm, Georg Albert PVD) and baked at 90 °C for 10 min, followed by immersion in acetone for 2 hours to remove the PMMA support.

*2.1.4 Transfer of graphene-CNM and $MoS_2$-CNM heterostructures*

For further characterization, graphene-CNM heterostructures as well as $MoS_2$-CNM heterostructures were transferred onto Si wafers with 300 nm of $SiO_2$ (Siltronix, roughness 0.3 nm RMS). The transfer was carried out in the same way as for the graphene on copper, however, a different etching solution ($I_2$/KI/$H_2O$ in mass proportion of 1:4:10) was used to remove the gold layer. The gold layer itself was separated from mica by a slight dipping into water of one of the edges/corners of the sample as described in detail in [21]. To obtain freestanding heterostructures, the same transfer protocol was applied to transfer the structures to TEM grids (copper grids, 400 mesh, different support films, Plano) with the exception that in this process to remove the PMMA support layer the acetone was exchanged with $CO_2$ in a critical point dryer (Autosamdri 815, Tousimis). This avoids ruptures in the samples, which otherwise might occur due to the surface tension of the evaporating solvent.

*2.1.5 Patterning of graphene*

To produce graphene patterns on gold/mica substrates, photolithography as well as electron beam lithography (EBL) was employed. For the photolithography a positive tone resist (AR-P 3510, Allresist, 2.0 μm) was exposed in a mask aligner (mercury arc lamp, Süss MicroTec) and developed in TMAH based solution (AR300-35, Allresist, 60 seconds). To obtain smaller feature sizes, a negative tone resist was used (AZ 5214 E, Microchemicals, 1.8 μm), exposed in a mask alignment system (EVG 620 TP, 30 mJ/cm²) and developed in a TMAH based solution (AZ 726 MIF, 45 seconds). For the smallest structures, a PMMA resist layer was spincoated on top of the samples, patterned by EBL (Vistec EBPG 5000plus) and subsequently developed. Pattern transfer to the graphene layer was achieved by reactive ion etching of the unprotected areas in an oxygen/argon plasma (Leybold Z401, Sentech Etchlab 200, 20-30 W, 30 seconds). The dissolution of the resist layer in acetone (2 hours) results in the fabrication of the desired graphene structures.

*2.1.6 Formation of self-assembled monolayers (SAMs)*

Structured graphene layers and $MoS_2$ crystals on gold substrates were introduced for 72 h to a ~1 mM solution of 4'-nitro-1,1'-biphenyl-4-thiol (NBPT, 99%, Taros) or biphenyl-4-thiol (BPT, 97%, Sigma Aldrich, individual graphene samples) dissolved in N,N-dimethylformamide (DMF, Alfa Aesar, 99.8%, water free) resulting in the formation of a SAM of aromatic thiols in the bare gold areas [24, 25]. After formation of the SAM, the samples were rinsed with DMF several times and blown dry in a nitrogen stream.

*2.1.7 Electron irradiation of graphene-SAM and $MoS_2$-SAM patterns*

When a SAM has been formed on the gold areas, the respective graphene-SAM and $MoS_2$-SAM samples were irradiated with electrons of 100 eV kinetic energy at a dose of 50 mC/cm² under high vacuum conditions ($1 \times 10^{-8}$ mbar). This converts the SAM into a CNM by an established electron induced crosslinking process [4, 26] and, as we demonstrate in this study, induces their stitching with the edges of graphene or $MoS_2$. This process is described in detail in the results section.

## 2.2  Microscopy characterization

### 2.2.1  Optical microscopy

The optical microscopy images were taken with a Zeiss Axio Imager Z1.m microscope equipped with a 5 megapixel CCD camera (AxioCam ICc5) in bright field operation. For optimal absorption contrast of the 2D materials, silicon substrates with 300 nm of oxide layer were used.

### 2.2.2  Scanning electron microscopy

The SEM images were taken with a Zeiss Sigma VP at a beam energy of 15 kV and use of the in-lens detector of the system.

### 2.2.3  Helium ion microscopy

The HIM images were taken with a Carl Zeiss Orion plus at 35 kV with 10 μm aperture, an integration time of 1 μs per pixel and line averaging with 64 lines. The HIM provides high material contrast on the 2D materials with very low charging effects.

### 2.2.4  Atomic force microscopy

The AFM measurements were performed with a Ntegra (NT-MDT) system in contact mode at ambient conditions using n-doped silicon cantilevers (CSG001, NT-MDT) with a typical tip radius of 6 nm and a typical force constant of 0.03 N/m.

*2.2.5 Transmission electron microscopy*

Transmission electron microscopy was carried out using an image-side corrected FEI Titan 80-300. To reduce the damage by electron irradiation, the samples were investigated at a low voltage of 80 kV. For enhancement of the image contrast and information limit, the extraction voltage of the Schottky-type electron source was lowered to 2000 V instead of the typical 4000 V [27]. The images were acquired using a Gatan Ultrascan 1000XP CCD camera with exposure times of 0.5 to 2.0 seconds with frame sizes of 2048×2048 pixels, and electron doses of about $5 \cdot 10^6$ $e^-/nm^2$. For more clear visualization between amorphous and crystalline regions in the HRTEM images, we Fourier-filter the HRTEM image and calculate an filtered image from the masked crystalline reflections, following the protocol described in [28]. In the filtered TEM image, similar to a dark-field image, the crystalline area appears bright and the amorphous area appears dark. Thus, the processed image easily allows to discriminate between regions consisting of CNM (amorphous) and of graphene (see Fig. 3b) or $MoS_2$,(see Fig. 8f). Electron energy-loss spectroscopy (EELS) and energy-filtered TEM (EFTEM) experiments on the $MoS_2$/CNM heterostructure were carried out with a Gatan GIF Quantum post-column energy filter. For the identification of the CNM and the $MoS_2$, we used the carbon K-edge (284 eV) and the sulfur L-edge (165 eV) signals, respectively. The shown C-K and S-$L_{2,3}$ elemental maps were obtained via the three-window background subtraction method,[29] with a 20 eV energy-selecting slit. Thickness maps showing the relative thickness of the membrane were calculated from zero-loss filtered and unfiltered TEM images using the log-ratio method.

*2.6 Spectroscopy characterization*

*2.6.1 Raman spectroscopy*

Most of the Raman spectra and mapping were acquired using a Bruker Senterra spectrometer operated in backscattering mode. Measurements at 532 nm were obtained with a frequency-doubled Nd:YAG Laser, a 50x objective and a thermoelectrically cooled CCD detector. The spectral resolution of the system is 2-3 $cm^{-1}$. Some spectra (see figure captions for details) were obtained using a Labram Aramis spectrometer at 473 nm with a 50x objective, a grating with 2400 lines/mm and a spectral resolution of 1 $cm^{-1}$. For all spectra, the Si peak at 520.7 $cm^{-1}$ was used for peak shift calibration of the instrument. The Raman spectroscopy maps were obtained using a motorized XY stage. For analysis of the characteristic $MoS_2$ and graphene peaks the background was subtracted and the data were fitted with Lorenzian functions using a LabVIEW script to determine the peak position, FWHM and maximum intensity of the peaks.

*2.6.2 X-ray photoelectron spectroscopy*

X-ray photoelectron spectra were taken in a Multiprobe UHV system (Omicron), using a monochromatic X-ray source (Al K$_\alpha$) and an electron analyzer (Sphera) with a resolution of 0.9 eV.

## 3. Results and discussions

*3.1 Engineering of lateral graphene-CNM and MoS$_2$-CNM heterostructures*

In Fig. 1 we schematically present the steps involved in the fabrication of graphene-CNM and MoS$_2$-CNM lateral heterostructures. First, graphene and MoS$_2$ sheets are grown by CVD on Cu foils or SiO$_2$/Si wafers, respectively, as described in Sections 2.1.1 and 2.1.2. After the growth they are transferred onto gold/mica samples. To induce patterns in the graphene sheets photolithography or electron beam lithography is employed (Section 2.1.5). As the MoS$_2$ monolayers (Section 2.1.2) form discontinuous films consisting of triangles with lateral dimensions of ~100 μm (Fig. S2b), no patterning was employed here. In the next step, the SAMs of aromatic biphenyl-thiols (BPT or NBPT, see Section 2.1.6) were grown on the gold areas free of graphene or MoS$_2$. Finally, to form the heterostructures, large area electron irradiation with an energy of 100 eV and a dose of 50 mC/cm$^2$ was employed (see Section 2.1.7). Upon this treatment, the SAMs are crosslinked forming the CNMs [4] whose edges are stitched with the edges of the graphene or MoS$_2$ sheets resulting in the formation of continuous 2D lateral heterostructures. The formed graphene-CNM and MoS$_2$-CNM heterostructures show significant mechanical stability and employing the polymer assisted transfer (see Section 2.1.4) can be transferred onto target substrates such as e.g. SiO$_2$/Si wafers or suspended on grids. In Sections 3.2 and 3.3 we present their detailed characterization by complementary microscopy and spectroscopy techniques.

*3.2 Characterization of lateral graphene-CNM heterostructures*

In Figs. 2a-b optical microscopy (OM) images of graphene-CNM heterostructures transferred onto a 300 nm SiO$_2$/Si wafer are shown. Fig. 2a presents a heterostructure with a regular pattern of graphene stripes with a width of 5 μm embedded in a CNM matrix and in Fig. 2b, a square array of 2.5×2.5 μm$^2$ graphene dots in a CNM matrix is presented. In the OM images, graphene areas appear darker, as they have a higher optical contrast in comparison to the surrounding CNM [30]. Both heterostructures are continuous within the whole microfabricated area, which corresponds to 10×10 mm² in total. Some rupture defects are visible in Figs. 2a-b enabling to differentiate clearly between the bare $SiO_2$ surface and the heterostructure areas. Note that some darker inclusions within the graphene patterns correspond to graphene double layers resulting from the CVD growth (see Fig. S1a). The subsequent preparation steps of the graphene-CNM heterostructures on gold/mica were characterized by helium ion microscopy (HIM). The pattern transfer into graphene sheets, the formation of SAMs on gold and finally the formation of graphene-CNM heterostructures can be followed by this technique from the respective changes in the contrast, (see Fig. S2). Next, we characterized the heterostructures after their transfer onto $SiO_2$ by atomic force microscopy (AFM). Fig. 2c shows a lateral force AFM image of the heterostructure presented in Fig. 2a. The chemically more homogeneous graphene areas reveal a lower contrast in the lateral force image in comparison to the CNM areas consisting of crosslinked biphenyl-thiols. The corresponding topography image (see Fig. S3) is in agreement with the thickness of both materials, which is 0.4 nm and 0.9 nm for graphene and BPT CNMs [18], respectively, and shows that the graphene areas have a lower height in comparison to the CNM areas. Some inhomogeneities are observed within CNM and graphene areas both in the topography and lateral force images, which most probably result from non-perfect adhesion of the heterostructure to the substrate and PMMA residuals. The HIM characterization of the samples on the growth gold/mica substrates and OM/AFM characterization of the samples transferred onto $SiO_2$/Si substrates demonstrate the electron-beam induced stitching of CNM and graphene and the formation of continues graphene-CNM heterostructures on a large scale.

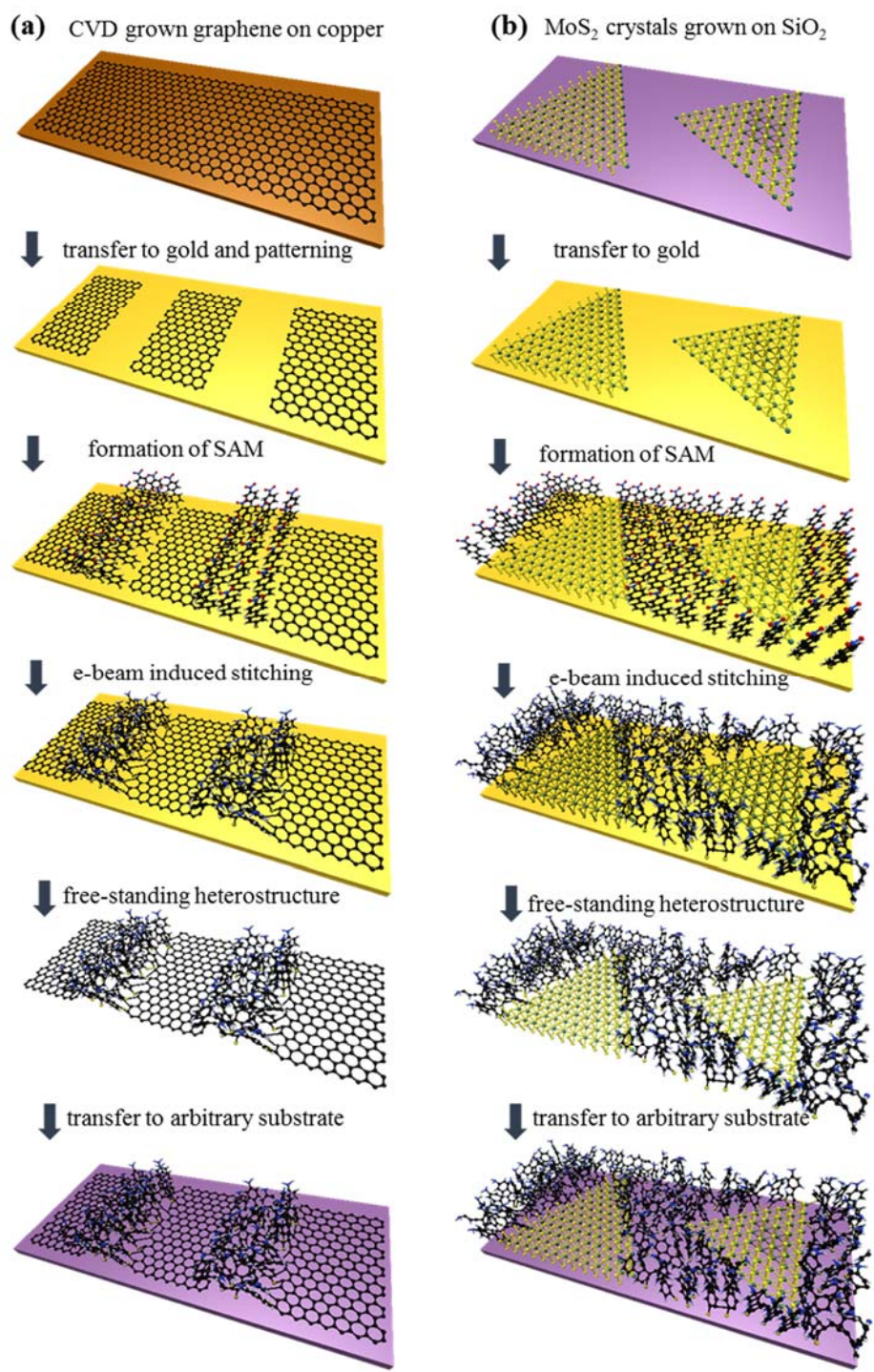

**Fig. 1.** (a) Scheme of the preparation of a lateral graphene-CNM heterostructure. A graphene layer grown on a copper foil is transferred to a gold substrate and lithographically patterned. Then a SAM of BPT or NBPT molecules (see Experimental for details) is grown in the bare gold areas and subsequently crosslinked by electron irradiation. This results in a continuous, heterogenous graphene-CNM layer, which then can be transferred to form a freestanding heterostructure or to an arbitrary substrate. (b) Scheme of the preparation of a lateral $MoS_2$-CNM heterostructure. Here, CVD grown monolayer crystals are used as $MoS_2$ structures. The other necessary steps are the same as in (a).

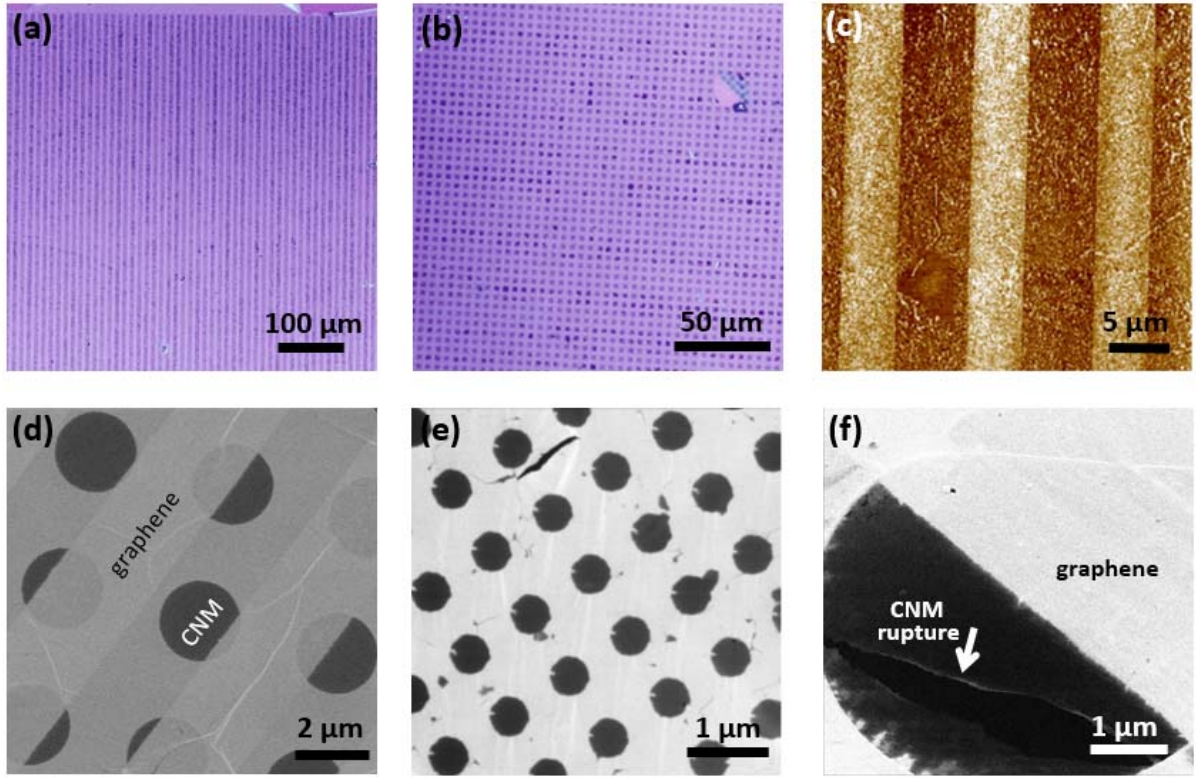

**Fig. 2.** Graphene-CNM heterostructures. (a-b) Optical microscopy (OM) images of lateral graphene-CNM heterostructures on a 300 nm SiO$_2$/Si substrate. The color are slightly adjusted to provide higher contrast. BPT SAM was used to form CNMs. Graphene appears darker than the CNM. The graphene stripes in (a) are 5 μm wide. The graphene square dots in (b) have a width of 2.5 μm. Only few defects are visible on a large scale. The whole sample has a size of 10×10 mm². (c) AFM lateral force image of the heterostructure presented in (a). The corresponding topography image is available in SI Fig. S3. (d) SEM image of a suspended graphene-CNM heterostructure with 2.5 μm wide stripes transferred to a Quantifoil TEM grid with circular holes. (e) HIM image of a graphene-CNM heterostructure suspended on a TEM grid. The CNM nanocircles of 500 nm in diameter are embedded in a large graphene monolayer. The hole pattern was induced in the graphene sheet by electron beam lithography (EBL). (f) HIM image showing the graphene-CNM heterostructure boundary. It is prepared freestanding on a holey carbon support film. The conducting graphene area has a higher secondary electron rate in comparison to insulating CNM and appears brighter in the image.[31] A ruptured CNM was chosen to provide better contrast between the two materials and free space. NBPT SAM was used to form CNMs in (d-f).

To characterize the mechanical robustness of the formed heterostructures, they were prepared as suspended sheets on TEM grids, Figs. 2d-f. Fig. 2d presents a SEM image of a similar graphene-CNM heterostructure as in Fig. 1a, which was transferred to a Quantifoil TEM grid with a square array of circular holes. In this image, the conductive graphene appears brighter than the insulating CNM. It can be seen that the suspended heterostructure preserves its mechanical integrity and has a sharp boundary between CNM and graphene. In Fig. 2e a HIM image of a suspended graphene-CNM heterostructures with a dot pattern is shown. This heterostructure was prepared by inducing a square array of circular holes into a graphene sheet. The dots with a diameter of 500 nm consist of CNM embedded into a matrix of graphene. This graphene-CNM heterostructure demonstrates an inverse material contrast in comparison to Fig. 2b, where a graphene dot pattern is embedded into a CNM matrix. Because of the higher secondary electron yield of graphene in comparison to CNM [4], similar to the SEM imaging (Fig. 2d), in HIM the bright areas correspond to graphene while the darker areas correspond to CNM. As shown in Fig. S4, suspended graphene-CNM heterostructures can be prepared on the millimeter scale with a low defect density. In our experiments, the maximum lateral size of the suspended heterostructures was restricted simply to the diameter of TEM grids (2.9 mm) with holey carbon film serving as a support for the suspended sheets. To characterize the grain boundary between CNM and graphene, high-resolution HIM imaging was employed, Fig. 2f. As seen from Fig. 2f, the width of the grain boundary, i.e. the stitching region between CNM and graphene appears to be very narrow, below the resolution of the applied microscopy technique. Therefore, aberration-corrected high-resolution transmission electron microscopy (AC-HRTEM) at 80 keV was employed to characterize this region in more detail.

The characterization of the boundary between CNM and graphene in a suspended heterostructure by AC-HRTEM is presented in Fig. 3. Fig. 3a shows an atomically resolved 30×30 nm$^2$ micrograph of the boundary region. Left to the drawn red line the typical graphene structure [18, 32] can be recognized, whereas to the right the disordered CNM [4] is imaged. In the filtered TEM image (Fig. 3b) from the AC-HRTEM image in Fig. 3a, the areas of crystalline graphene (bright contrast) and amorphous CNM (dark contrast) are easily visualized; the boundary is also marked in the AC-HRTEM image by the red line. The interface between graphene and CNM is straight within an accuracy of few nm, and appears to be molecularly sharp. It is known that the formation of a CNM from aromatic molecular precursors via electron irradiation includes the cleavage of carbon-hydrogen bonds and the crosslinking of the adjacent molecular backbones [26]. Since the edges of graphene sheets are terminated with hydrogen of other chemical groups, a similar crosslinking mechanism including cleavage of the terminal chemical groups and formation of new covalent bonds between carbon atoms in graphene and CNM most probably takes place during the electron irradiation induced stitching.

During the subsequent production steps of graphene-CNM heterostructures, we characterized the graphene by Raman spectroscopy. The corresponding spectra are presented in Fig. 4a. For the samples measured on gold/mica substrates the background intensity, due to the metal substrate, was subtracted from the respective spectra. The lowest spectrum in Fig. 4a represents a typical graphene sample transferred onto a gold/mica substrate. The characteristic 2D- and G-peaks at 2680 cm$^{-1}$ and 1593 cm$^{-1}$ respectively, have an intensity ratio of I(2D)/I(G)=2.0; in the wave number range typical for the D-peak (~1350 cm$^{-1}$) some intensity increase is observed, which is difficult to quantify due to the noisy background. For similar graphene samples studied on SiO2/Si substrates, a typical I(D)/I(G)-ratio of 0.1 was observed indicating a low defect density in the graphene (Fig. S1c) [33]. After patterning the graphene into an array with the line width of 5 μm, the I(2D)/I(G)-ratio decreases to 1.1 and the position of the 2D-peak shifts to the higher wave numbers (2700 cm$^{-1}$). This observation is indicative for a change in the electron density in graphene [33] caused most probably by the doping of the graphene at the edges or/and on the graphene surface resulting from the processing by photolithography (EBL) and reactive ion etching (RIE) (see Section 2.1.5). The red line in Fig. 4a represents the signal from an area where the graphene was removed by RIE; here no characteristic 2D- and G-peaks are detected and the spectral shape near the D-peak is similar to that observed for the non-patterned graphene on gold. Next, after the formation of a BPT SAM on the graphene free gold areas no additional changes are observed in the graphene spectra indicating that the structure and the electron density remain unaffected. As seen from the magenta spectrum, the SAM itself does not show any Raman signal in the studied spectral range. By crosslinking the SAM into a CNM with electron irradiation and formation of the graphene-CNM heterostructure, the D-peak appears in the spectrum. Finally, after the transfer of this heterostructure onto SiO$_2$ substrate, because of the much lower background intensity, the D, G and 2D peaks can be analyzed in more detail (see the upper spectra in Fig. 4a). Thus we find that the I(2D)/I(G)-ratio decreases to 0.8 and the I(D)/I(G)-ratio has a value of 0.3 demonstrating an increased defect density in the graphene. Fig. 4b shows an OM image and the respective Raman maps of the graphene-CNM heterostructure on SiO$_2$. Raman active graphene areas and the non-active CNM areas can be clearly distinguished. In the CNM areas, the Raman signals of defective graphene occasionally are observed, most probably resulting from some graphene residuals after its removal by RIE.

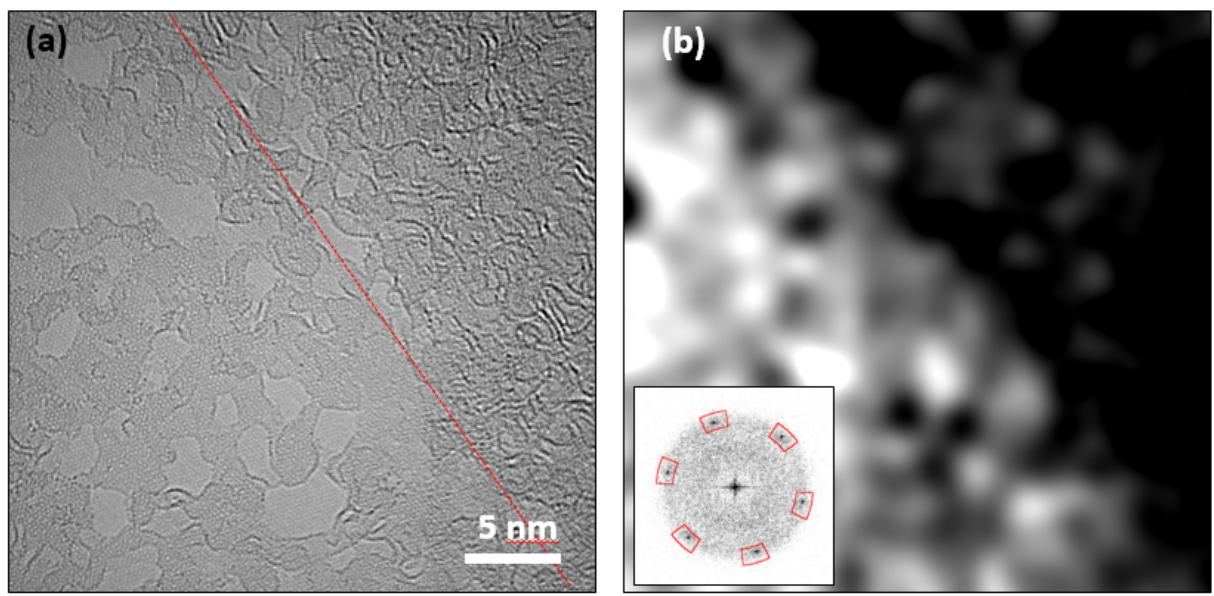

Fig. 3. (a) AC-HRTEM image showing the boundary between CNM (top right) and graphene (bottom left) marked by the dotted red line. (b) Filtered image of (a) showing the crystalline graphene area by bright contrast and the amorphous CNM area by dark contrast. The inset in (b) shows the Fourier Transform of (a) with the masked graphene reflection for the filtered image in (b).

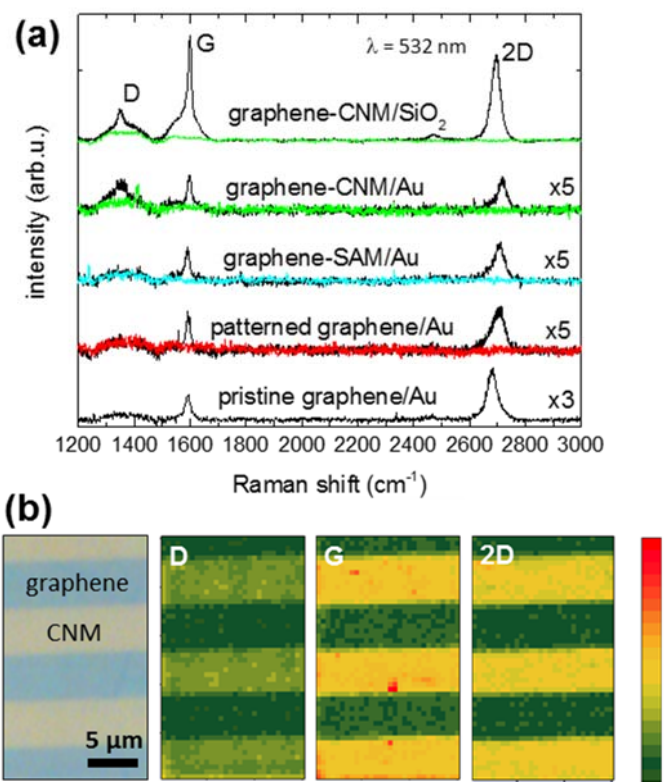

**Fig. 4.** (a) Raman spectra of graphene (black), an area where the graphene has been removed by reactive ion etching (red), a SAM (blue) and the CNM (green) on gold substrates (lower spectra) and on $SiO_2$ (uppermost spectrum), respectively. For the spectra taken on gold samples, a background subtraction was performed. (b) Optical microscopy image of a lateral graphene-CNM heterostructure (left). The graphene stripes appear darker than the CNM. The same area was chosen for a Raman mapping (right); the intensities of the graphene D, G and 2D peaks are presented.

Next, we studied the effect of the electron irradiation on graphene by complementary Raman and X-ray photoelectron spectroscopy (XPS) measurements, Fig. 5. To this end, graphene sheets from the same CVD growth were transferred onto SiO$_2$ substrates and exposed to electron irradiation ($E_{kin}$=50 eV) in ultra-high vacuum (UHV, 2×10$^{-9}$ mbar) with electron doses of 10 mC/cm², 25 mC/cm² and 50 mC/cm². After the irradiation they were characterized by Raman spectroscopy at ambient conditions and subsequently by XPS in UHV. As seen from Fig. 5a, an increase of the D-peak is substantial only for the sample irradiated with 25 mC/cm$^2$, whereas for the non-irradiated sample and for the samples irradiated with 10 mC/cm$^2$ and 50 mC/cm$^2$ its intensity is comparable. The respective XP spectra show that the core level electron C1s-signal for the 25 mC/cm2 sample has the widest full width at half maximum (FWHM) in comparison to the other samples, Fig. 5b. This observation is indicative for the presence of a higher surface density of the polymeric contaminations on this sample resulting from the PMMA based transfer. Thus, we attribute the observed increase in the D-peak to the appearance of new defects in the graphene via the formation of new carbon-carbon bonds with the hydrocarbons adsorbed on its surface [34]. Besides that, the I(2D)/I(G)-ratio in all samples continuously decreases upon the irradiation from 3.0 to about 0.8 and the positions of the G-peak and the 2D-peak shifts by 16 cm$^{-1}$ to higher wavenumbers. These changes in the Raman spectra can be due to an increase of the p-doping in graphene [34]. A similar effect was reported for graphene samples annealed in UHV on SiO$_2$ and exposed to ambient conditions. Here, the observed p-doping was attributed to the physisorbtion of molecular adsorbates on the graphene surface [35]. Summarizing these findings, we conclude that upon electron irradiation induced stitching of graphene and CNMs the modification of the structural and electronic properties of graphene can take place. On the one hand, defects can be introduced into graphene if some hydrocarbons are present on its surface; on the other hand, an additional p-doping at ambient condition is induced most probably due to the physisorbtion of molecular adsorbates.

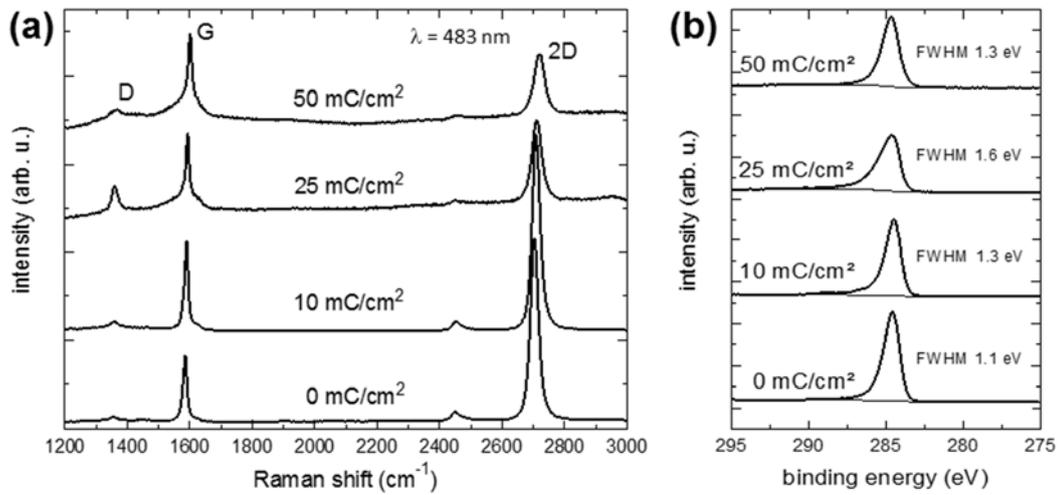

**Fig. 5.** (a) Raman spectra of graphene samples transferred to SiO$_2$ and irradiated with different electron doses at a beam energy of 50 eV. (b) X-ray photoelectron spectra of the C1s core electrons of the graphene samples described in (a). The full width at half maximum (FWHM) of the respective peaks is included.

*3.3 Characterization of lateral CNM-MoS$_2$ heterostructures*

In this section we present the characterization of lateral MoS$_2$-CNM heterostructures formed by the procedure described in Section 3.1. An OM image of a MoS$_2$-CNM heterostructure transferred onto a SiO$_2$ substrate is shown in Fig. 6a. The NBPT precursor (see Section 2.1.6) was used for the generation of CNM, resulting in its termination with the amino groups [17]. In Fig. 6a the CNM, MoS$_2$ and SiO$_2$ regions can be clearly distinguished from the optical contrasts. The total area of this heterostructure is about 1×1 cm$^2$. To prove the lateral stitching between MoS$_2$ and CNM, we transferred the heterostructure onto a holey Quantifoil TEM grid and imaged by HIM, Fig. 6b. Due to the higher secondary electron yield from semiconducting MoS$_2$ in comparison to insulating CNM, the MoS$_2$ area appears much brighter in this image than the CNM area. In the central part of Fig. 6b, in the suspended region, a hole in the CNM is visible. This defect helps to identify the free-standing boundary between MoS$_2$ and CNM, that is, to confirm the successful stitching between both materials. The ability to transfer the MoS$_2$-CNM heterostructures from their growth substrates onto new substrates as well as the preparation of the suspended sheets demonstrates their high mechanical robustness during the microfabrication.

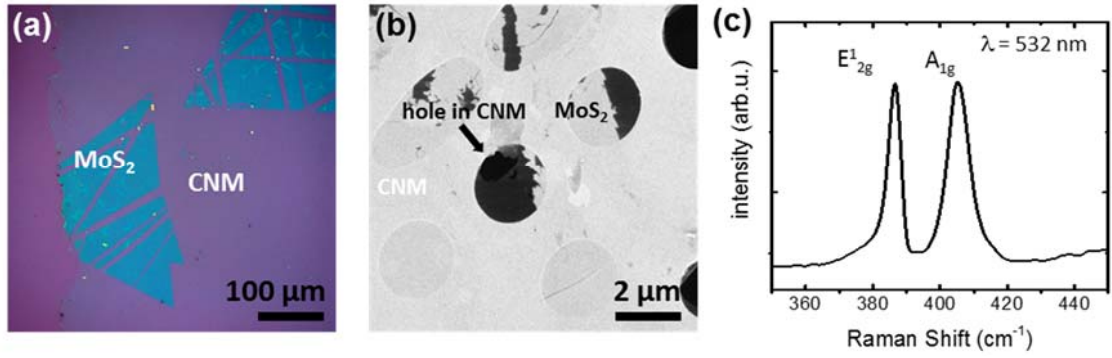

**Fig. 6.** (a) Light microscopy image of a lateral MoS$_2$-CNM heterostructure on a 300 nm SiO$_2$/Si substrate. (b) HIM image of a freestanding MoS$_2$-CNM heterostructure transferred to a Quantifoil TEM grid. (c) Raman spectrum recorded from the MoS$_2$ region of the heterostructure after transfer to the SiO$_2$.

Next, we employed Raman spectroscopy to analyze the influence of the electron irradiation, used for the stitching, on the structure of MoS2. Fig. 6c shows a Raman spectrum from the MoS2 area of the heterostructure transferred on SiO2 (see Fig. 6a). The spectrum reveals the characteristic for a MoS2 monolayer peaks at 386.8 ± 0.2 cm-1 (FWHM = 4.5 ± 0.2) and 404.9 ± 0.2 cm-1 (FWHM = 6.8 ± 0.3), which are due to the in-plane (E12g band) and out-of-plane (A1g band) vibrations in the sheet [36, 37], respectively. Within the experimental errors we do not find any noticeable change in this spectrum in comparison to the spectrum of as grown MoS2 on SiO2 (see Fig. S1d and SI Table 1) with an exception that the FWHM of A1g peakincreases from $4.9 \pm 0.1$ cm$^{-1}$ for as grown MoS$_2$ to $6.8 \pm 0.3$ cm$^{-1}$ for the MoS$_2$ in the heterostructure which is indicative for a change in the doping.[36] Thus the results suggest that no significant structural modification in the MoS$_2$ monolayer occurs upon the electron irradiation.

The MoS$_2$-CNM heterostructures on SiO$_2$ were characterized further by AFM in the contact mode. We have found that in topography imaging the MoS$_2$ region appears about 2-3 nm higher than the CNM region, Fig. S5a. Its roughness has the root mean square (RMS) value of 1.7 nm, which is significantly higher than the RMS value of 0.3 nm typically observed for the as grown MoS$_2$ monolayers on SiO$_2$. Moreover, in disagreement with the expectation, only a weak contrast between the chemically and tribologically heterogeneous MoS$_2$ and CNM regions of the heterostructure was observed in the lateral force image, Fig. S5b. These findings strongly indicate that an additional organic layer has been formed on top of the MoS$_2$ during the heterostructure fabrication. This layer may result from physisorbed or/and covalently bound (e.g., to sulfur vacancies in the MoS$_2$ [38-40]) NBPT molecules resulting from the SAM preparation. To prove this hypothesis, we performed AFM scans of the heterostructure with a slightly higher force than applied for imaging. In Figs. 7a-b an 8×8 μm² area of the MoS$_2$-CNM heterostructure is shown, which was scanned in this way. As seen from the topography image in Fig. 7a, the adsorbate layer was removed by the AFM tip from the MoS$_2$ surface revealing its typical topographic features like e.g. folds in this atomically thin sheet. The height difference between the MoS$_2$ and CNM regions becomes to be negligible. On contrast, in the lateral force image, Fig. 7b, the MoS$_2$ area becomes clearly distinguishable from the laterally stitched CNM. The SEM imaging of this area presented in Fig. 7c and Fig. S6 shows that the MoS$_2$ region with the removed organic layer has a higher contrast and appears darker in comparison to the surrounding. Note that after removal of the adsorbed layer, no changes are observed in the area covered by CNM both in AFM topography and lateral force imaging demonstrating that the CNM remains unaffected by the scan with a higher force.

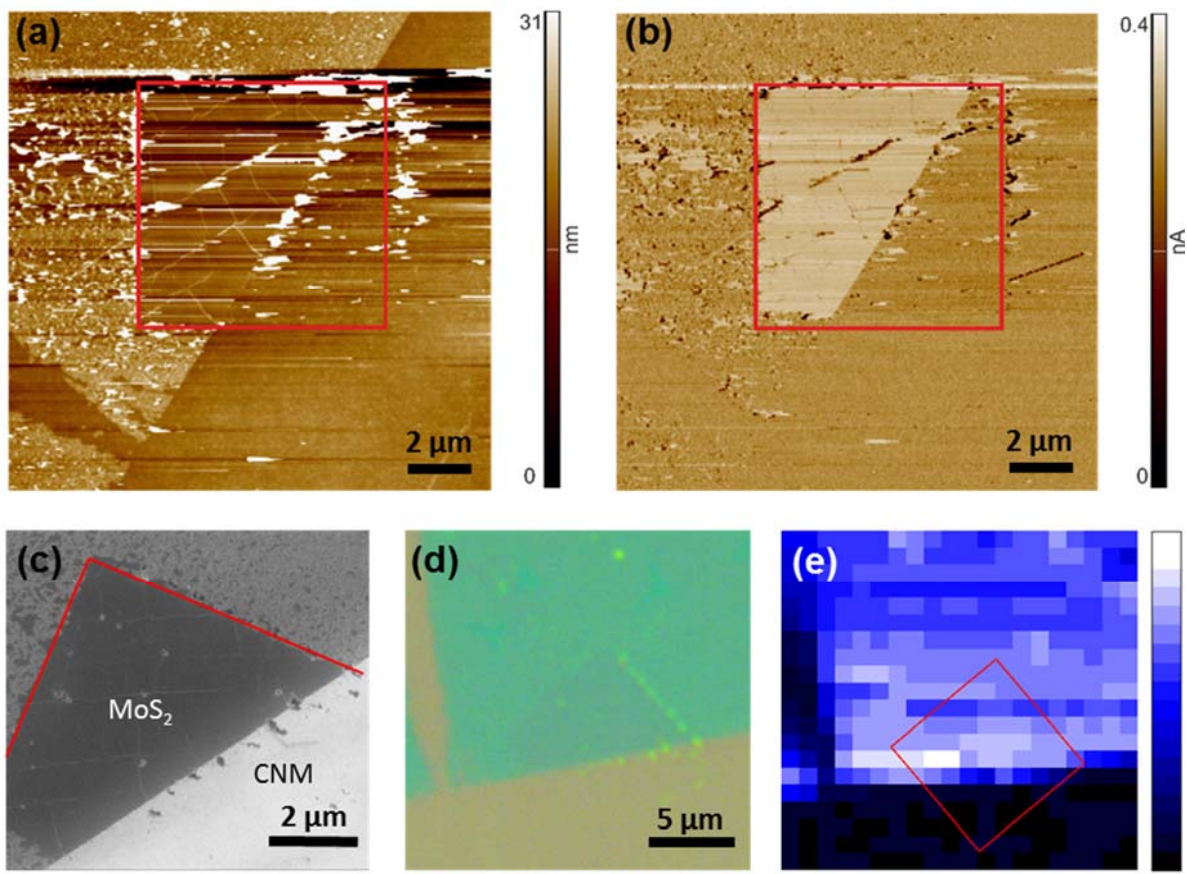

**Fig. 7.** AFM topography (a) and lateral force (b) image of a $MoS_2$-CNM heterostructure on a $SiO_2$ substrate. With increased force, the adsorbate weakly bound to the $MoS_2$ could be removed (red box). The AFM tip did not induce any damage to the underlying $MoS_2$ crystal or to the CNM layer laterally attached to it. (c) SEM image of the region shown in (a). (d) Light microscopy image of the heterostructure after the molecule removal with AFM. False colors are used to provide a higher contrast, the region where the adsorbate is removed can be recognized. (e) A Raman map of the investigated $MoS_2$ edge shows a uniform intensity distribution for the $E^1_{2g}$ mode in the area covered with $MoS_2$ and no signal in the area covered with CNM, proving that neither the electron irradiation nor the removal of weakly bound molecules damages the $MoS_2$ layer.

A detailed analysis of the SEM images (Figs. 7c and S6) shows that the adsorbed layer scanned by AFM with the imaging parameters, that is applying a low force, reveals some dark patches which are most probably due to the adsorbate removal. This observation indicates that the adsorbate layer is weakly bound to the MoS2 by physisorption. In order to verify this, we performed Raman spectroscopy and mapped the region where the adsorbed layer was removed by AFM as described in the previous paragraph, Figs. 7d-e and Fig. S7. This region in Figs. 7a-c,e is marked with red squares. From the intensity of the Raman $E^1_{2g}$ mode in Fig. 7d and the respective OM image in Fig. 7e the MoS$_2$ area can clearly be recognized. The statistical analysis of the obtained data (see SI Table 1) shows no differences in the spectral features inside and outside of the MoS$_2$ area where the molecules were removed by AFM, which demonstrates that the structure of the MoS$_2$ remains unaffected after the adsorbate removal. We conclude that the organic layer on top of the MoS$_2$ is physisorbed and most likely consists of the NBPT molecules employed for the SAM growth. Note that the formation of this layer can be prevented in microfabrication by protecting the MoS$_2$ areas with a polymeric film during the SAM growth on gold.

Similar as for to the graphene-CNM heterostructures, the AFM, HIM and SEM data suggest the formation of sharp boundaries between MoS$_2$ and CNM in the heterostructures. To study their structure on the nanoscale, HRTEM and chemical mapping by energy-filtered TEM (EFTEM) were applied. A TEM image and the respective carbon and sulfur maps of a free-standing MoS$_2$-CNM heterostructure transferred on a circular hole of the Quanifoil grid are presented in Fig. 8a-c. The uniform carbon content seen in Fig. 8b (C-K map) indicates the presence of a molecular carbonaceous layer on top of MoS$_2$. As expected, the strongest carbon signal in the C-K map can be found on the Quantifoil film, due to its much higher thickness of about 10 nm, compared to the 1 nm thin heterostructure. From the sulfur map in Fig. 8c (S-L$_{2,3}$ map), the MoS$_2$ region in the heterostructure is clearly recognized. An AC-HRTEM image and the respective Fourier transformation obtained from the MoS$_2$ are presented in Fig. 8d. They demonstrate the crystalline atomic structure of the MoS$_2$ monolayer; some disordered bright and dark fringes in the HRTEM image result most probably from the physisorbed molecular carbon layer. An AC-HRTEM image of a boundary between MoS$_2$ and CNM is shown in Fig. 8e. In the filtered image in Fig. 8f, the areas of crystalline MoS$_2$ (bright contrast) and amorphous CNM (dark contrast) are visualized (applying the same procedure as described for Fig. 3.) The stitching region recognized from

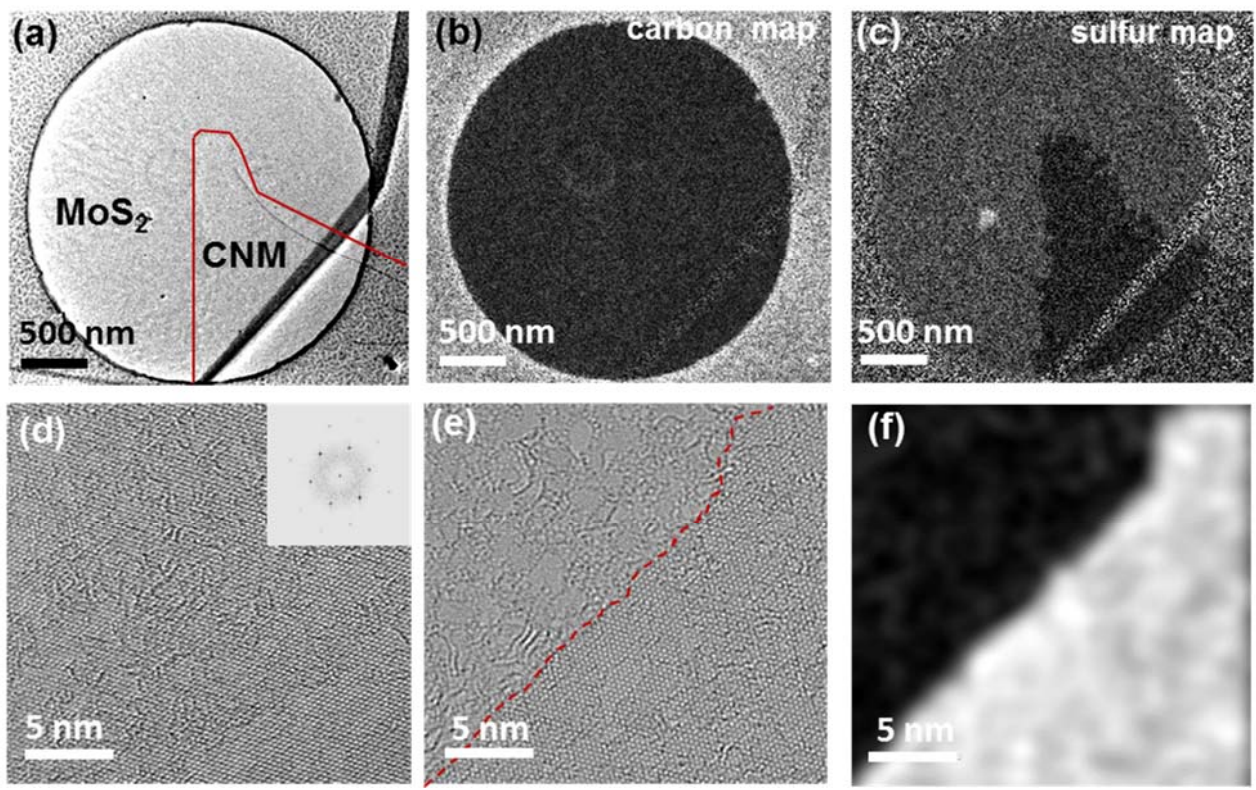

**Fig. 8.** (a) Overview TEM image of a $MoS_2$-CNM heterostructure on a Quantifoil grid with a freestanding circular part of 2 μm diameter. The boundary of the $MoS_2$ crystal is marked with a red line. (b) The EFTEM carbon map (C-K edge) of the same sample area shows a homogeneous carbon distribution on the free-standing heterostructure area showing the presence of an organic adsorbate on the $MoS_2$. (c) A sulfur map (S-$L_{2,3}$ edge) shows the high sulfur content in the region composed of $MoS_2$. (d) AC-HRTEM image and Fourier transform (inset) of the image obtained from a region with $MoS_2$. (e) AC-HRTEM image showing CNM (top left) and MoS2 (bottom right) and the boundary between the two regions marked red. (f) Filtered image of (e) showing the crystalline MoS2 area (bright contrast) and the amorphous CNM area (dark contrast).

Figs. 8e-f appears to be nearly atomically sharp. We assume that the sulfur atoms in the CNM bind to the molybdenum atoms and/or fill the sulfur vacancies at the edge of the $MoS_2$ [38-40] during the self-assembly and the electron irradiation; in parallel with the crosslinking of the aromatic cores in the SAM, this process results in the formation of a continuous $MoS_2$-CNM heterostructure with a nearly atomically sharp boundary between these two materials.

## 4. Summary

We have presented a methodology for the synthesis of lateral heterostructures of 2D materials by the low electron energy irradiation induced stitching. The heterostructures composed of inorganic-organic atomically thin sheets - graphene-CNM and $MoS_2$-CNM - were produced and characterized by a combined microscopy and spectroscopy study down to the nanoscale. Our results demonstrate that the formed heterostructures possess a high mechanical robustness, so that they can be transferred from the growth substrates onto new solid substrates or prepared and suspended sheets without mechanical damage. Moreover, the formed grain boundaries between the dissimilar materials appear to be nearly atomically sharp with a width below 2 nm. Our results show that the electron irradiation doses required for the stitching do not noticeably influence the pristine properties of the used graphene or $MoS_2$ monolayers. Employing conventional lithography techniques, the heterostructures can be produced in any shape, which facilitates their potential for applications in electronic and optoelectronic devices. As the irradiation with low energy electrons can induce chemical reactions in a variety of materials (see e.g. [41]), we suggest that the developed methodology can potentially be applied to a broad family 2D organic and inorganic materials paving the way towards flexible synthesis of their lateral heterostructures with complex architectures incorporating into the atomically thin sheets some distinct electronic, optical and biochemical functions.

**Acknowledgements**

We acknowledge financial support of the Deutsche Forschungsgemeinschaft (DFG) through SPP 1459 "Graphene" (TU149/2-2, KA1295/19-2), Heisenberg Programm (TU149/3-1), research grant TU149/5-1 and research infrastructure grant INST 275/257-1 FUGG. "ProExzellenz 2014-2019" grant of the Thüringer MWWDG. MJM, JB and UK further acknowledge the DFG and the Ministry of Science, Research and the Arts (MWK) of Baden-Wuerttemberg in the frame of the SALVE (Sub Angstrom Low-Voltage Electron microscopy) project (KA1295/21-1). This project has also received funding from the European Union's

Horizon 2020 research and innovation programme under grant agreement No 696656. We thank Stephanie Höppener and Ulrich S. Schubert for the access to Raman spectroscopy and SEM. Armin Gölzhäuser is acknowledged for the possibility to conduct HIM measurements.

**Appendix A. Supplementary data**

Supplementary data related to this article can be found

https://doi.org/10.1016/j.carbon.2017.11.034